\documentclass[12pt]{iopart}

\usepackage{iopams}
\usepackage{graphicx}
\usepackage{psfrag}
\usepackage{mathrsfs} 

\newcommand{\pd}[2]{\frac{\partial #1}{\partial #2}}

\newcommand{\norm}[1]{\left\| #1 \right\|}
\newcommand{\Scri}{${\mathscr I}^+$}

\begin{document}

\title{Tails for the Einstein-Yang-Mills system}

\author{Michael P\"urrer and Peter C Aichelburg}
\address{Institut f\"ur Theoretische Physik,
         Universit\"at Wien,
         1090 Wien, Austria}
\ead{Michael.Puerrer@univie.ac.at, Peter.Christian.Aichelburg@univie.ac.at}

\begin{abstract}
  We study numerically the late-time behaviour of the coupled Einstein Yang-Mills system.
  We restrict ourselves to spherical symmetry and employ Bondi-like coordinates with
  radial compactification. Numerical results exhibit tails with exponents close to
  $-4$ at timelike infinity $i^+$ and $-2$ at future null infinity \Scri.
\end{abstract}


\section{Introduction} 
\label{sec:introduction}

Radiating systems relax to equilibrium by dissipating energy to infinity. 
The fall-off properties of the field at late times are governed by so-called 
radiation tails. These tails emerge from primary outgoing radiation that is 
backscattered. This far-field effect is either due to a background or an effective
potential produced by the nonlinearities of the radiation field itself, or in general, 
a mixture of both.

The classical fall-off properties were based on liner perturbations about a given background
\cite{Price72}. However, Bizon \cite{Bizon-tails-general-08} has pointed out, based on previous
results \cite{asymptotic-stability-skyrmion07,szpak-tails-I,szpak-tails-II}, that for certain
nonlinear systems the nonlinear tails may dominate the long-time behaviour, i.e. these tails 
fall off more slowly in time than the linear perturbations. As an example, he studied the 
spherically symmetric Yang-Mills equations on Minkowski and Schwarzschild spacetimes
\cite{bizon-tails-YM-Mink-SS}. While linear perturbation theory predicts a $t^{-5}$ power law decay
due the backscattering off the Schwarzschild curvature, the nonlinear part decays only as $t^{-4}$
for observers near timelike infinity. This slower fall-off was also observed numerically. More
recently, these results were confirmed and extended to the late-time behaviour at future null
infinity by Zengino\u{g}lu \cite{Anil-tails}, showing that tails die off as $t^{-2}$ on 
Schwarzschild spacetime.

These calculations are on a given background and the question arises what
happens for the coupled Einstein-Yang-Mills system. In the fully coupled
case it is difficult to disentangle the different contributions. If one
writes down a  perturbation expansion 
starting from flat spacetime, then the first order perturbation of the YM field 
evolves in an effective potential produced by the back reaction of the
YM-field  to the metric.  In addition, there will be the nonlinear
effects from the YM equation already present in flat space. In this sense
all tails are nonlinear, however, it is not evident what the overall fall-off 
behaviour is. The aim of this paper is to answer this question numerically.

It is known that the possible stable endstates of (spherically symmetric)
collapse for the Einstein-YM system are either the formation of a black
hole or dispersion to flat space. Black holes come in two types:  pure black holes,
i.e. with all of the YM field radiated away or coloured \cite{bizon-colored-bhs}. 
In addition, there exist soliton solutions found by Barnik and McKinnon \cite{Bartnik-McKinnon}. 
However, both solitons and colored black holes turned out to
be unstable under linear perturbations. For technical reasons we restrict ourselves to
subcritical evolutions which do not form black holes. However, from the
results in \cite{bizon-tails-YM-Mink-SS,Anil-tails}, we expect our findings to hold also for 
generic spherically symmetric initial data when a black hole is formed.

We tackle this problem by numerically solving a characteristic initial value problem
with constraints and employ radial compactification to allow investigation of tails
at future null infinity.
First we present the model, derive the equations of motion and write the Yang-Mills
wave equation as an advection equation plus a constraint.
Second, we discuss the numerical methods used for solving the given system and check
the convergence of the code.
Finally, we present results for late-time tails on Minkowski background and for the 
coupled Einstein-Yang-Mills system.


\section{Model} 
\label{sec:model}

We assume spherical symmetry with a regular center and choose Bondi-like coordinates 
$\{u, r, \theta, \phi \}$ based upon outgoing null hypersurfaces $u = const$ with the line-element \cite{Puerrer-Husa-PCA}
\begin{equation}
  ds^2 = -e^{2\beta(u,r)} \frac{V(u,r)}{r} du^2 - 2 e^{2\beta(u,r)} du dr + r^2 d\Omega^2.
\end{equation}

We consider the Yang-Mills theory with the gauge group SU(2) and assume
the magnetic ansatz for the gauge connection \cite{bizon-tails-YM-Mink-SS,Choptuik-Hirschmann-Marsa-EYM}
\begin{equation}
  A = w \tau^\theta d\theta + \left( \cot\theta \tau^r + w \tau^\phi \right) \sin\theta d\phi,
\end{equation}
where $w = w(u,r)$ is the Yang-Mills field and the $\tau^a$ are the spherical
generators of $su(2)$, normalized such that $[\tau^a,\tau^b] = i \epsilon^{abc} \tau^c$,
where $a,b,c \in \{ r,\theta,\phi \}$.
They are related to the Pauli matrices $\sigma^a$ via $\tau^a = \sigma^a / 2$.

The Yang-Mills field strength (or curvature), $F = dA + A \wedge A$, then becomes
\begin{eqnarray}
  F = \left( \dot w du + w' dr  \right) \wedge \left( \tau^\theta d\theta + \tau^\phi \sin\theta d\phi \right)
    - \left( 1 - w^2 \right) \tau^r d\theta \wedge \sin\theta d\phi,
\end{eqnarray}
where $\dot w$ and $w'$ denote partial derivatives of $w(u,r)$ with respect to $u$
and $r$, respectively.
Using the above ansatz the trace of the Yang-Mills curvature becomes 
\begin{equation}
    \tr \left(F^{\mu\nu}F_{\mu\nu}\right) = F^a_{\mu\nu} F^{a\mu\nu} = 
    - 2 \frac{e^{-2\beta}}{r^2} \left[ 2 \dot w w' - \frac{V}{r}(w')^2 \right] 
    +   \frac{(1-w^2)^2}{r^4},
\end{equation}
where Greek indices range over the four spacetime dimensions and Latin indices are group
indices.
The action for the Yang-Mills field coupled to Einstein's equations is \cite{Choptuik-Hirschmann-Marsa-EYM}
\begin{equation}
  S = \int d^4 x \sqrt{-g} \left[ \frac{R}{16\pi G} - \frac{1}{e^2} F^a_{\mu\nu} F^{a\mu\nu} \right],
\end{equation}

The Yang-Mills wave equation is obtained by varying the action with respect to the Yang-Mills 
field $w$
\begin{equation}\label{eq:YM-matter}
  -2 \dot w' + \left(\frac{V}{r}\right)' w' + \frac{V}{r} w'' + \frac{e^{2\beta}}{r^2} w(1-w^2) = 0,
\end{equation}
while variation with respect to the metric functions, $\beta$ and $V$, yields 
two constraint equations
\begin{eqnarray}
  V'     &= e^{2\beta}\left[ 1 - \frac{8\pi G}{e^2} \frac{(1-w^2)^2}{r^2} \right]\\
  \beta' &= \frac{8\pi G}{e^2} \frac{(w')^2}{r}.
\end{eqnarray}

The coupling constant $[G/e^2]$ has dimension of $length^2$. Since it is not dimensionless, changing the coupling constant does not give rise to a one-parameter 
family of theories, but only changes the scale.
To simplify the equations, we choose 
\begin{equation}
  \frac{8\pi G}{e^2} = 1.
\end{equation}
The final form of the hypersurface equations then becomes
\begin{eqnarray}
  \label{eq:geometry-V-eq}
  V'     &= e^{2\beta}\left[ 1 - \frac{(1-w^2)^2}{r^2} \right]\\
  \label{eq:geometry-beta-eq}
  \beta' &= \frac{(w')^2}{r}.
\end{eqnarray}

The regularity condition to be imposed on the Yang-Mills field at the origin is
\begin{equation}\label{eq-w-regularity}
  w = \pm 1 + \Or(r^2)
\end{equation}
while the gauge ($u$ is chosen to be proper time at the regular center) 
and regularity conditions on the metric functions are
\begin{eqnarray}
  \beta &= \Or(r^2)\\  
  V/r &= 1 + \Or(r^2).
\end{eqnarray}


For the study of tail behaviour it is crucial to introduce a new field variable
\begin{equation}
  \bar w := w - 1  
\end{equation}
as a perturbation of one of the Yang-Mills vacua $w = \pm 1$, (i.e. where the field 
strength vanishes), in order to avoid the tails being swamped by accumulated
numerical errors.
Since the matter field equation and the constraint equations are invariant under reflection
symmetry, $w(u,r) \to \tilde w(u,r) := - w(u,r)$, we may specialize to one of the two vacua
without loss of generality.

There are a number of different ways to go about solving the Yang-Mills wave equation
\eref{eq:YM-matter} in Bondi coordinates, e.g. the diamond integral approach due to 
Gomez and Winicour 
\cite{Gomez92a,Winicour98} which we used in \cite{Puerrer-Husa-PCA} or Goldwirth and 
Piran \cite{Goldwirth-Piran-mKG} and Garfinkle's \cite{Garfinkle95} way of rewriting 
the equation with a total time derivative along the ingoing null geodesics. 
The latter allows for employing standard \emph{method of lines} (MOL) techniques, 
i.e. one first discretizes the spatial derivatives (albeit non-equispaced), 
which in turn, yields a system of coupled ordinary differential equations (ODEs) 
that can be solved using a standard ODE solver. 
In this approach to solving the characteristic initial value problem, the gridpoints 
usually move along the ingoing null geodesics, which implies a nontrivial origin treatment, 
with Taylor expansions for increased accuracy.

In contrast to the methods mentioned above, we prefer to simply introduce a new evolution
variable 
\begin{equation}
  h := \bar w_{,r}  
\end{equation}
so as to eliminate the mixed $ur$ derivative in equation \eref{eq:YM-matter}. 
In addition, we also keep the locations of gridpoints (in time) 
at fixed values of $r$. Here, MOL discretizations, using standard stencils for
equidistant grids, are applicable and the treatment of the origin is trivial 
modulo boundary conditions.
Moreover, we are here not interested in strong field regions, where the focussing of
ingoing null geodesics would provide a natural increase of resolution near the center. 
Rather, we want to study late time tails for subcritical evolutions, which entails 
tracking the field at locations of constant $r$ through time.

The evolution equation then becomes
\begin{equation}\label{eq:YM-matter-MOL}
  \dot h = \frac{1}{2} \left(\frac{V}{r}\right)' h + \frac{1}{2}\frac{V}{r} h' 
  - \frac{1}{2} e^{2 \beta} \frac{F(\bar h)}{r^2},
\end{equation}
where
\begin{equation}
  F(\bar h) = 2 \bar h + 3 \bar h^2 + \bar h^3,
\end{equation}
and $\bar h \equiv \bar w$.

We now have an added constraint to solve (in addition to the two geometry equations \eref{eq:geometry-V-eq}, \eref{eq:geometry-beta-eq})
\begin{equation}\label{eq:h-bar-constraint}
  \bar h = \int_0^r h(u,\tilde r) d \tilde r.
\end{equation}

Note that equation \eref{eq:YM-matter-MOL} is of advection type. For flat space and
without the YM self-interaction term, it reduces to 
\begin{equation}\label{eq:advection-flat-ingoing}
  \dot h = \frac{1}{2} h',
\end{equation}
which is equivalent to the flat space wave equation for 
\begin{equation}\label{eq:wave-flat-constraint}
  \phi = \frac{1}{r} \int_0^r h(u,\tilde r) d \tilde r.
\end{equation}
Given initial data $h(u=0,r) = f(r)$ with $r \in \mathbb{R}$ the solution of the
advection equation \eref{eq:advection-flat-ingoing} is simply
\begin{equation}
  h(u,r) = f(r + \frac{1}{2} u), \quad r \in \mathbb{R}, \, u > 0.
\end{equation}
The characteristic curves $r + u/2 = const$ are purely ingoing and thus, there is no 
boundary condition at the origin of spherical symmetry. The outgoing characteristic
of the wave equation comes into play via the constraint equation \eref{eq:wave-flat-constraint}.

The outer boundary needs a different treatment.
While it is possible to specify an outgoing wave boundary condition we prefer to 
use compactification, which has the advantage of being able to observe late-time behaviour 
at future null infinity.
As in \cite{Puerrer-Husa-PCA} we introduce a compactified 
radial coordinate 
\begin{equation}
  x := \frac{r}{1+r},
\end{equation}
which maps $r \in [0,\infty] \mapsto x \in [0,1]$.
In addition, we use the Misner-Sharp mass-function
\begin{equation}
  m = \frac{r}{2}\left[ 1 - g^{rr} \right] 
    = \frac{r}{2}\left[ 1 - \frac{V}{r} e^{-2\beta} \right]
\end{equation}
as an evolution variable, thereby eliminating $V$.
This is necessary, since the compactified constraint equation for $V$ is singular at future 
null infinity \Scri, similar to the scalar field case treated in \cite{Puerrer-Husa-PCA}.
Instead of $h$ we introduce a new field variable
\begin{equation}
  \tilde h := \bar w_{,x} = \bar h_{,x},
\end{equation}
which finally leads to a manifestly non-singular evolution system.

The evolution equation then becomes
\begin{equation}\label{eq:YM-matter-MOL-cmp}
  {\tilde h}_u = 
  \frac{1}{2} \left[e^{2 \beta} \left(1 - 2m\frac{1-x}{x}\right) (1-x)^2 \tilde h\right]_{,x}
  -\frac{1}{2} e^{2 \beta} \frac{F(\bar h)}{x^2},
\end{equation}
and the constraints are
\begin{eqnarray}
  \label{eq:bar-h-constraint-cmp}
  \bar h &= \int_0^x \tilde h d\tilde x\\
  \label{eq:beta-constraint-cmp}
  \beta_{,x} &=  \frac{(1-x)^3}{x} (\tilde h)^2\\
  \label{eq:m-constraint-cmp}
  m_{,x} &= \left[1 - 2m \frac{1-x}{x}\right](1-x)^2 (\tilde h)^2
  + \frac{\bar h^2 (4 + 4 \bar h + \bar h^2)}{2x^2}
\end{eqnarray}

The regularity conditions at the origin for the compactified scheme become, using $r = \Or(x)$,
\begin{eqnarray}
  \tilde h &=  \Or(x) \qquad  \;\,{\bar h}&= \Or(x^2)\\
  \beta &= \Or(x^2)   \qquad  m &= \Or(x^3)
\end{eqnarray}
There is no boundary condition at future null infinity, since it is a characteristic.


\section{Numerics} 
\label{sec:numerics}

From a numerical point of view, we do not even need to know that we are solving a
characteristic initial value problem. Rather, we may simply take the
evolution system and solve the nonlinear advection equation \eref{eq:YM-matter-MOL-cmp} 
using a standard MOL approach in conjunction with computing the constraint ODEs
\eref{eq:bar-h-constraint-cmp}, \eref{eq:beta-constraint-cmp}, and \eref{eq:m-constraint-cmp}.

We may discretize the advection term in \eref{eq:YM-matter-MOL-cmp} 
by centered, fully upwind or upwind-biased stencils. Combined with an ODE integrator for 
time, such as, the classical 4th order Runge-Kutta method (RK4), the resulting 
schemes will then exhibit different numerical errors. One has to make a choice between 
added dispersion, in the case of centered approximations, and added dissipation for upwind schemes. 
These errors affect mostly high frequency components of the solution and can be minimized by 
using high order approximations. 

In the interior of the grid we have chosen a 6th order upwind-biased scheme with the stencil
\begin{equation}\label{eq:6th-order0upwind-biased-stencil}
\fl  \pd{H}{x}(x_i) = \frac{2H_{i-2} - 24 H_{i-1} - 35 H_i + 80 H_{i+1} - 30 H_{i+2} + 8 H_{i+3} - H_{i+4}}{60 \Delta x} 
+ \Or(\Delta x^6),
\end{equation}
where $H_i = H(x_i)$.
The above upwind stencil is the one closest to the centered stencil and it also has the
lowest error term among the upwind stencils \cite{sascha-6th-order-fd}.
The two alternate upwind stencils do not lead to stable evolutions in our case.
The weights for such finite-difference stencils can conveniently be computed in a 
computer algebra package, such as Mathematica, using Fornberg's compact algorithm
\cite{Fornberg-FD-weights}.
In contrast to centered schemes, where it is often necessary to add some artificial dissipation
to have numerical stability, the dissipation is already ``built-in'' in our chosen scheme.

The term $F(\bar h) / x^2$ in \eref{eq:YM-matter-MOL-cmp} forces a regularity 
boundary condition at the origin, so that
\begin{equation}
  \tilde h(u,x=0) = 0.
\end{equation}
We enforce it by choosing initial data that satisfy this condition (to machine precision)
and then make sure that its time derivative is zero, i.e. $\tilde h_u(u,x=0) = 0$ for all timesteps.
We choose Gaussian  initial data 
\begin{equation}\label{eq:initial-data}
  \bar h(0,x) = A \exp \left[ -200 (x - 1/2)^2 \right].   
\end{equation}

The constraint equations for $\bar h$ and $\beta$ are integrated using cumulative
Newton-Cotes quadrature rules of order 6, which are given in the appendix.
Since the right hand side of the constraint equation for the Misner-Sharp mass-function 
\eref{eq:m-constraint-cmp}
depends on the unknown $m$, we integrate it with RK4 and use 4th order polynomial
interpolation for computing $\tilde h$ and $\bar h$ at points $x_{i+1/2}$ in between actual
gridpoints, as required by the Runge-Kutta method.

In comparison to the established methods for solving characteristic initial value problems
\cite{Gomez92a,Winicour98, Goldwirth-Piran-mKG, Garfinkle95}, the method used here has a number 
of advantages for the problem at hand. It relies on a standard MOL discretization using 
equidistant stencils and thus allows for the use of high order schemes. Moreover, the boundary treatment is simple and does not require Taylor series near the origin.
It also allows us to track the field at lines of constant $r$ without further interpolation.

\begin{figure}[htbp]
  \centering
  \psfrag{u}{$\scriptstyle u$}
  \psfrag{Cf}{$\scriptstyle C_f$}
  \includegraphics[width=0.75\textwidth]{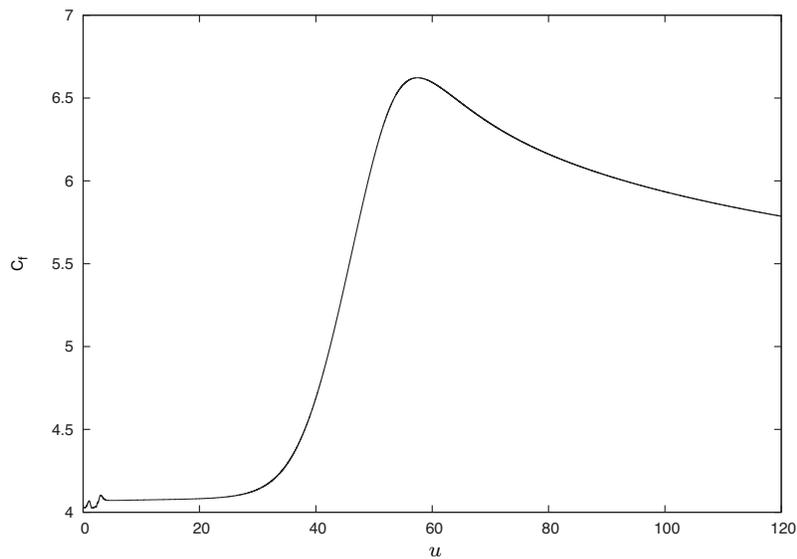}
  \caption{This figure shows the convergence 
  factor  $C_f = \log_2 \frac{ \norm{\bar h^{1000} - \bar h^{500}} }
                      { \norm{\bar h^{2000} - \bar h^{1000}} }$
  in the $l2$-norm of the solution for the coupled EYM-system with grid resolutions of 500, 
  1000 and 2000 gridpoints and a Courant number of $C = 1.25$. 
  Initially the code is 4th order convergent. Since the initial amplitude $A = 0.28$ is quite 
  large and $2m/r \approx 0.6$ the spatial accuracy is only 4th order.
  Later in the evolution, as the field disperses, $2m/r$ becomes very small and the spatial
  discretization becomes roughly 6th order accurate. At even later times the convergence
  order decreases slowly - probably due to the standard degrading of convergence in long 
  evolutions.}
  \label{fig:EYM-6Up-conv-milne-uf=80-C=0.99-high-1600-amp=0}
\end{figure}

In general, we expect the code to be 4th order convergent, see figure
\ref{fig:EYM-6Up-conv-milne-uf=80-C=0.99-high-1600-amp=0}.
For small data, however, $\max 2m/r$ is also small.
Since $m$ only appears in the wave equation in this combination, errors in the computation 
of $m$ may be allowed to be larger than those in other fields, and we may therefore have 
close to 6th order spatial accuracy, in this case. 
If the Courant number 
\begin{equation}
  C := \frac{1}{2} \frac{\Delta u}{\Delta x}
\end{equation}
is small enough, so that the errors from the RK4 integrator 
are of order $\Or(\Delta x^6)$, then, we may in fact achieve 6th order convergence.
It is, however, impractical to have small Courant numbers for very long time evolutions.
Clearly, evolutions using high Courant numbers complete faster and need less timesteps than
evolutions using lower Courant numbers. Therefore, accumulation errors should also be somewhat less
severe for high Courant numbers. Moreover, the classical RK4 solver exhibits 
additional damping \cite{Durran-wave-eq-num} of high frequency modes near the stability limit, 
which is beneficial to numerical stability.

For weak fields, the Yang-Mills advection equation \eref{eq:YM-matter-MOL-cmp} 
approximately reduces to
\begin{equation}
  {\tilde h}_u = \frac{1}{2} (1-x)^2 \tilde h_x.
\end{equation} 
After freezing the nonconstant coefficient $(1-x)^2$ at its maximum $1$,
Fourier analysis leads to the stability limit (similar to the 6th order centered case discussed
in \cite{Gustafsson-Kreiss-Oliger})
\begin{equation}
  C = \frac{1}{2} \frac{\Delta u}{\Delta x} \leq 1.29
\end{equation}
for a MOL discretization with the 6th order upwind-biased stencil
\eref{eq:6th-order0upwind-biased-stencil} and RK4 as time-integrator.
For the coupled system, we have found that Courant numbers of up to $C \leq 1.25$ yield
stable evolutions.

We have used numerical Python to conveniently automate the determination of tail 
exponents via fitting, while the core of the code was written in C++.


\section{Results} 
\label{sec:results}

\begin{figure}[htbp]
  \centering
  \includegraphics[height=3in]{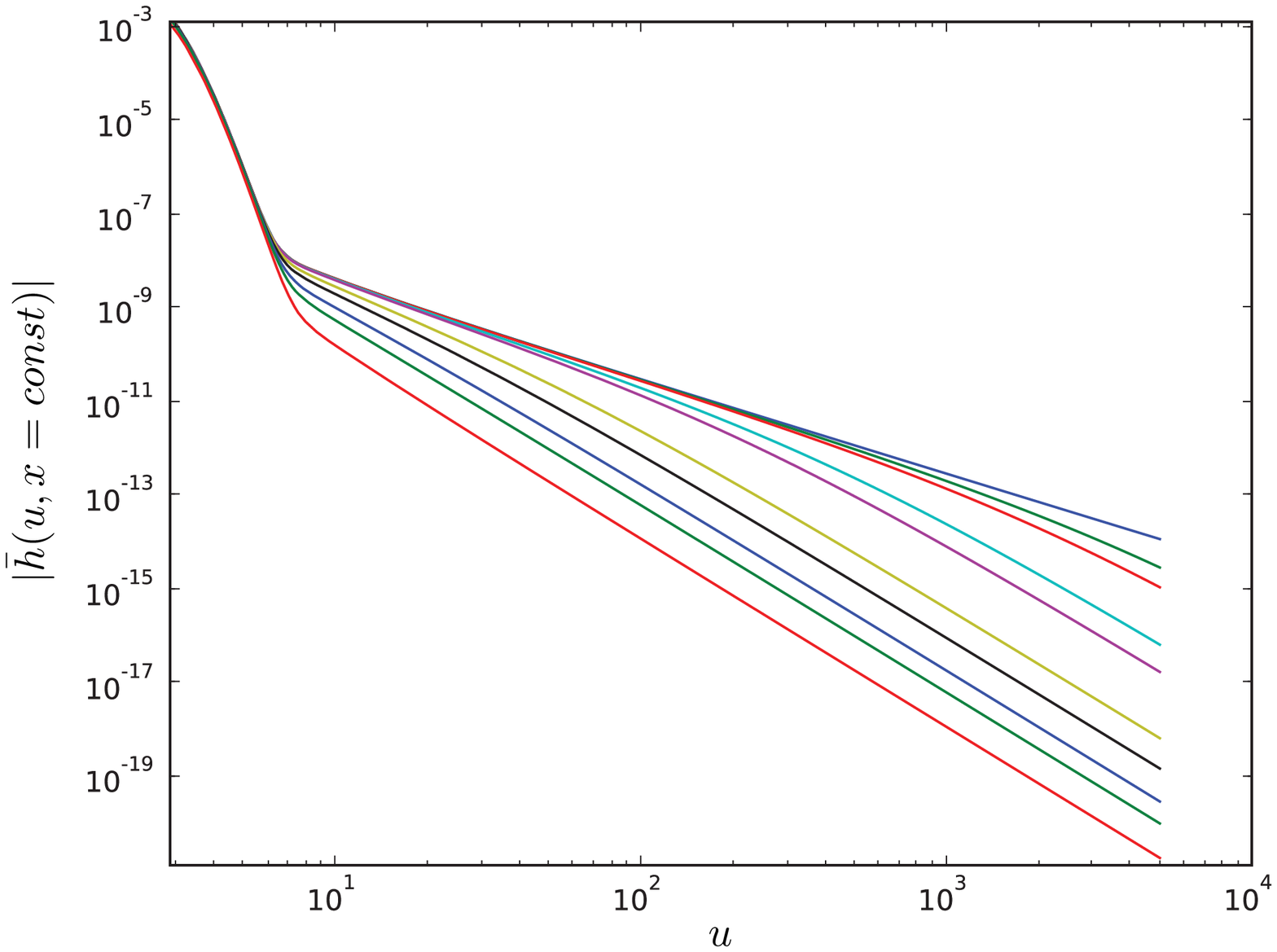}
  \psfrag{Tail exponents}[b]{$\qquad \quad \scriptstyle p(u,x)$}
  \psfrag{u}{$\scriptstyle u$}
  \psfrag{Tail Exponents from Fits}{}
  \includegraphics[height=3in]{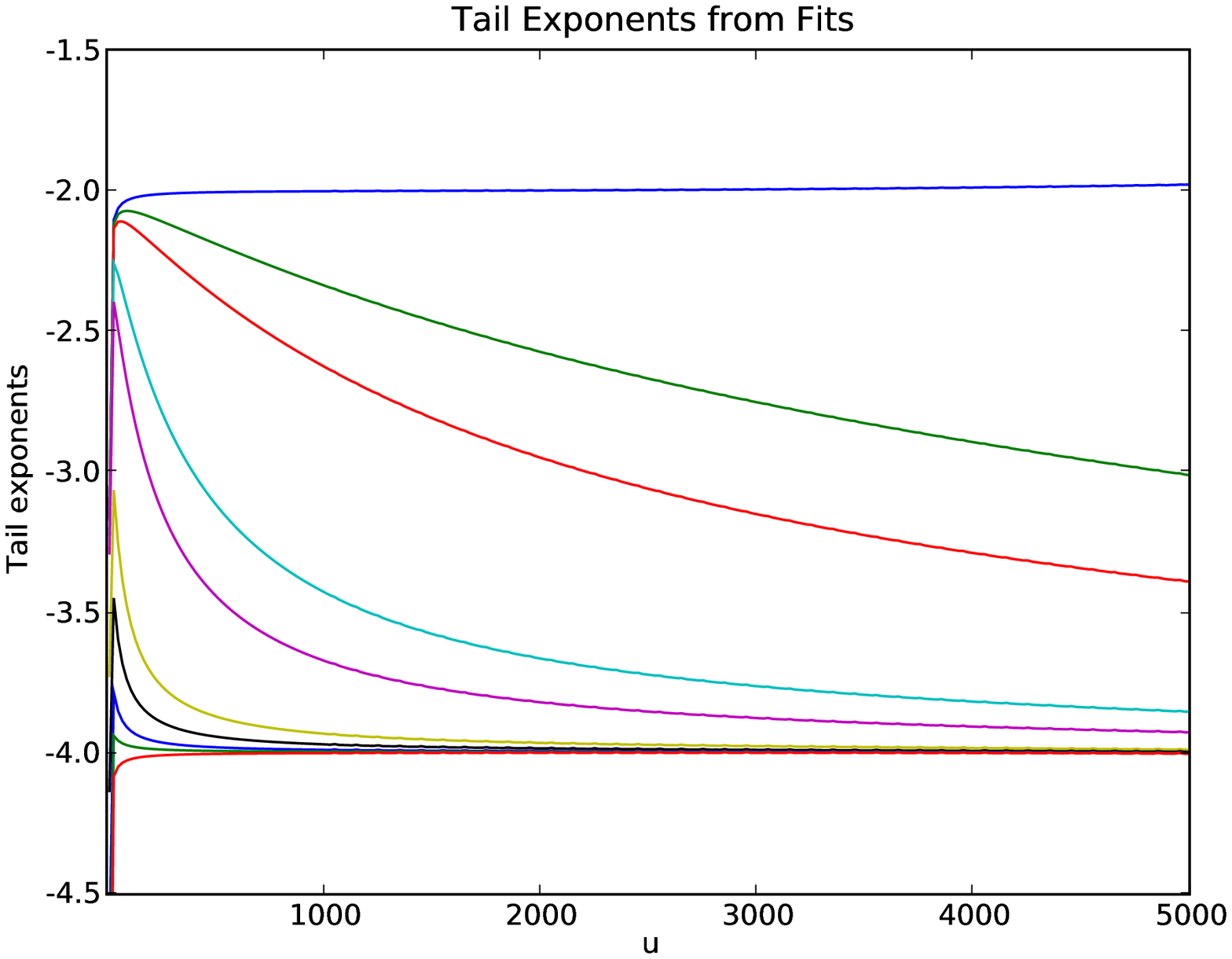}
  \caption{
    The upper plot shows the decay of the field $\bar h$ at $x=const$ versus retarded time 
    $u$, while the lower plot depicts the respective tail exponents for the same evolution 
    on Minkowski background. On \Scri the tail exponent is close to $p=-2$. Observers located
    at finite $x$ (or $r$) approach timelike infinity $i^+$ with the exponent $p=-4$ for 
    late times. Observers closer to the center approach this value faster than those closer
    to \Scri.
  }
  \label{fig:YM-FLAT-6Up-milne-tails-uf=5000-C=2_5-10000gp-amp=0_01}
\end{figure}

Perturbation theory predicts the late-time behaviour of the solution to be
\begin{equation}
 \bar h(u,x) \sim C u^{p},
\end{equation}
where $p$ is the tail exponent.
A first test case for our code is to correctly reproduce the know tail behaviour on 
Minkowski background \cite{bizon-tails-YM-Mink-SS}.
In figure \ref{fig:YM-FLAT-6Up-milne-tails-uf=5000-C=2_5-10000gp-amp=0_01}
we find that the tail exponent tends to $p=-4$ for observers near timelike infinity $i^+$ 
and the exponent $p=-2$ for future null infinity \Scri. This also corresponds to the decay
found on Schwarzschild backgrounds \cite{Anil-tails}.
In terms of the compactified radial coordinate $x$, the observers are located at 
$x=(1, 0.9995, 0.999, 0.995, 0.99, 0.95, 0.9, 0.8, 0.7, 0.5)$.
For the results presented here, we have used $10000$ spatial gridpoints, a Courant number
of $C=1.25$ and an initial data amplitude of $A=0.01$. 
For bigger, but still subcritical (in the coupled case), amplitudes and/or smaller Courant
numbers, the tail decay is essentially the same.


\begin{figure}[htbp]
  \centering
  \includegraphics[height=3in]{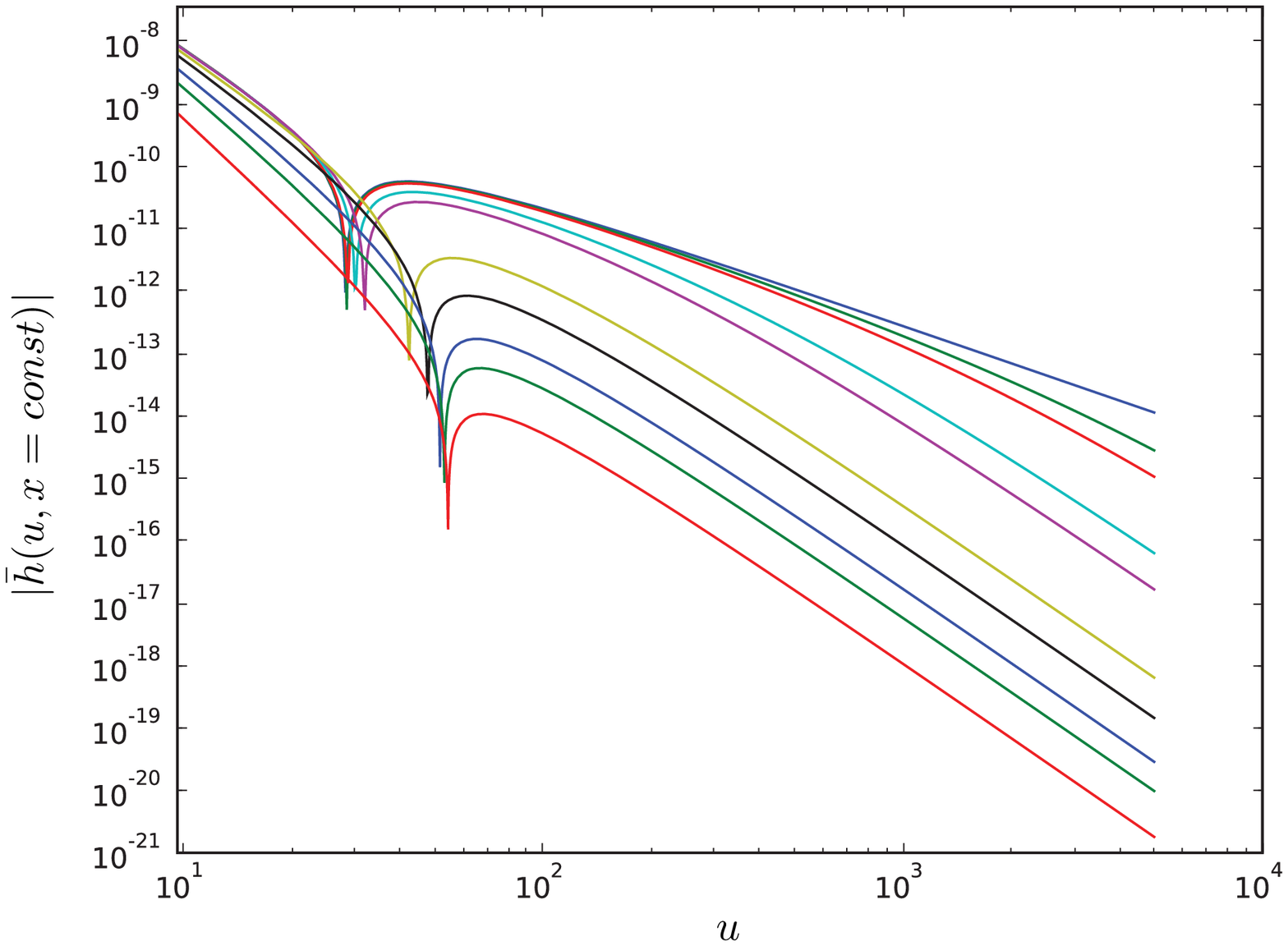}
  \psfrag{Tail exponents}[b]{$\qquad \quad \scriptstyle p(u,x)$}
  \psfrag{u}{$\scriptstyle u$}
  \psfrag{Tail Exponents from Fits}{}
  \includegraphics[height=3in]{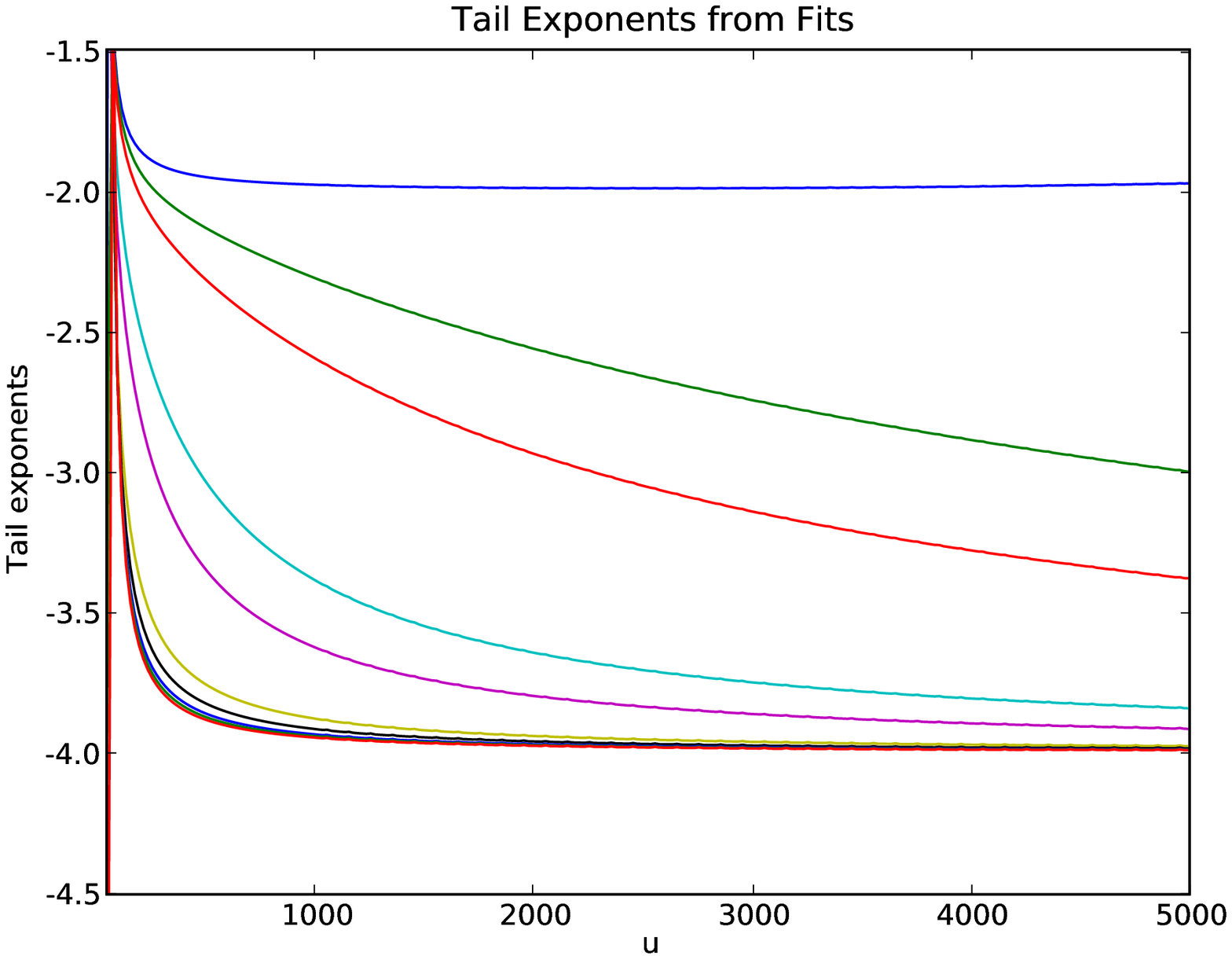}
  \caption{
    Similar to figure \ref{fig:YM-FLAT-6Up-milne-tails-uf=5000-C=2_5-10000gp-amp=0_01} these plots depict the late-time tails and the respective
    exponents for the coupled Einstein-Yang-Mills system. The exponents coincide with the
    behaviour on Minkowski space, being $p=-2$ at \Scri and $p=-4$ at $i^+$, respectively.
    The zero-crossing in $\bar h$ at $u \approx 50$ depends on the initial data amplitude,
    i.e. the field goes through zero earlier for smaller amplitudes.
  }
  \label{fig:EYM-6Up-milne-tails-uf=5000-C=2_5-10000gp-amp=0_01}
\end{figure}

Figure \ref{fig:EYM-6Up-milne-tails-uf=5000-C=2_5-10000gp-amp=0_01} encodes our main results showing
the late-time behavior for the coupled Einstein-Yang-Mills system. We find essentially the same
fall-off as for the Yang-Mills field on Minkowski background. As mentioned in the introduction, in
terms of a perturbative approach, tails are generated on the one hand by the the nonlinearity of the
YM field itself, and, on the other by the contribution of the field to the metric. What we see
numerically is a superposition of both effects which we can not separate. 
Hod\cite{Hod-tails} has studied linear wave tails in time dependent potentials. 
He finds that for a certain class of potentials that go to zero asymptotically in time, the
fall-off behaviour of the tails is a power law depending on the time dependence of the potential.


\section{Conclusion} 
\label{sec:conclusion}

Using Bondi-like coordinates and radial compactification we have written the Einstein-Yang-Mills system as an advection equation plus three constraints. In this form, MOL discretizations are
straightforward to apply, the origin treatment is easy and the outer boundary, \Scri, being a
characteristic does not require boundary data.
Compared to the diamond integral scheme \cite{Gomez92a,Winicour98} and Goldwirth, 
Piran and Garfinkle's  \cite{Goldwirth-Piran-mKG,Garfinkle95} way of solving 
characteristic initial value problems, the approach used here is very clean, simple to 
implement and allows the use of high order schemes.

We have found that the spherically symmetric coupled Einstein-Yang-Mills system shows the same
fall-off behaviour at late times as Yang-Mills on Minkowski or Schwarzschild backgrounds. Although
such a result could have been expected it is by no means evident, because so far it is not known 
how tails arising from the back reaction of the the Yang-Mills field to the metric decay. 
Our results indicate that they decay as fast or faster than the nonlinear tails on Minkowski
background.

Our compactified code has allowed us to also study the fall-off behavior at future null infinity. As
for the coupled Einstein massless scalar field \cite{Puerrer-Husa-PCA}, the fall-off on \Scri is
slower than for observers approaching timelike infinity. Since realistic observers are located only
at finite distances from the center, what then is the practical relevance to know the decay
conditions on \Scri? It has been pointed out in \cite{Puerrer-Husa-PCA} and also in
\cite{Anil-tails}, that for astrophysical observers, the relevant decay rate is the one along null
infinity. This has to do with the observation that the tail exponents for observers far out start
close to the exponent on \Scri and only slowly decrease to the value for timelike observers.


\ack
We thank Piotr Bizon for helpful discussions and comments on the manuscript. This work has been supported by the Austrian Fonds zur F\"orderung der wissenschaftlichen Forschung (FWF) (project P19126-PHY). Partial support by the Fundaci\'on Federico and the hospitality at the
Mittag-Leffler Institute (Sweden) is also acknowledged.

\appendix
\section*{Appendix}\label{sec:appendix}
\setcounter{section}{1}

The quadrature formulas below have been obtained by simply integrating the (quartic)
interpolating polynomial $P(f|x_{i-1},x_i, x_{i+1}, x_{i+2}, x_{i+3})$ on an equidistant 
grid with spacing $h$ over the intervals $[x_{i-1}, x_i]$ until $[x_{i+2}, x_{i+3}]$,
respectively.

\begin{eqnarray}
\fl  \int_{x_{i-1}}^{x_i} f dx  &= \frac{h}{720} \left(251 f_{i-1} + 646 f_i - 264 f_{i+1} + 106 f_{i+2} - 19 f_{i+3} \right) + \Or(h^7)\\
\fl  \int_{x_i}^{x_{i+1}} f dx  &= \frac{h}{720} \left(-19 f_{i-1} + 346 f_i + 456 f_{i+1} - 74 f_{i+2} + 11 f_{i+3} \right) + \Or(h^7)\\
\fl  \int_{x_{i+1}}^{x_{i+2}} f dx  &= \frac{h}{720} \left(11 f_{i-1} - 74 f_i + 456 f_{i+1} + 346 f_{i+2} -  19 f_{i+3} \right) + \Or(h^7)\\
\fl  \int_{x_{i+2}}^{x_{i+3}} f dx  &= \frac{h}{720} \left(-19 f_{i-1} + 106 f_i - 264 f_{i+1} + 646 f_{i+2} + 251 f_{i+3} \right) + \Or(h^7)
\end{eqnarray}
Summing these formulas together, yields the classical Boole's or Milne's rule.
\begin{equation}
\fl  \int_{x_{i-1}}^{x_{i+3}} f dx = \frac{2h}{45} \left(7 f_{i-1} + 32 f_i + 12 f_{i+1} + 32 f_{i+2} + 7 f_{i+3} \right)  + \Or(h^7)
\end{equation}

   
\bibliographystyle{unsrt}
\bibliography{EYM_tails}

\end{document}